\documentclass[prl,showpacs,twocolumn,superscriptaddress]{revtex4}
\usepackage{amsmath,amsfonts}
\usepackage{graphicx}% Include figure files
\usepackage{dcolumn}

\begin{document}

%=============    Title   ==================================
\title{Room temperature Peierls distortion in small diameter nanotubes}

\author{D. Conn\'etable}
\affiliation{
	Laboratoire de Physique de la Mati\`ere Condens\'ee
        et des Nanostructures (LPMCN),\\
	CNRS and Universit\'e Claude Bernard Lyon I,\\
	B\^atiment Brillouin, 43 Bd du 11 Novembre 1918,
        69622 Villeurbanne Cedex, France.
}
\author{G.-M. Rignanese}
\affiliation{
       Unit\'e de Physico-Chimie et de Physique des Mat\'eriaux (PCPM),
       Universit\'e Catholique de Louvain,
       1 Place Croix du Sud,
       B-1348 Louvain-la-Neuve, Belgium 
}
\affiliation{
       Research Center on Microscopic and Nanoscopic Electronic
       Devices and Materials (CERMIN),\\
       Universit\'e Catholique de Louvain,
       B-1348 Louvain-la-Neuve, Belgium
}
\author{J.-C. Charlier}
\affiliation{
       Unit\'e de Physico-Chimie et de Physique des Mat\'eriaux (PCPM),
       Universit\'e Catholique de Louvain,
       1 Place Croix du Sud,
       B-1348 Louvain-la-Neuve, Belgium 
}
\affiliation{
       Research Center on Microscopic and Nanoscopic Electronic
       Devices and Materials (CERMIN),\\
       Universit\'e Catholique de Louvain,
       B-1348 Louvain-la-Neuve, Belgium
}
\author{X. Blase}
\affiliation{
	Laboratoire de Physique de la Mati\`ere Condens\'ee
        et des Nanostructures (LPMCN),\\
	CNRS and Universit\'e Claude Bernard Lyon I,\\
	B\^atiment Brillouin, 43 Bd du 11 Novembre 1918,
        69622 Villeurbanne Cedex, France.
}

\date{\today}

%=============== ABSTRACT ==========================================
\begin{abstract}

By means of {\it ab initio} simulations, we investigate the phonon band
structure and electron-phonon coupling in small 4-\AA\ diameter nanotubes.
We show that both the C(5,0) and C(3,3) tubes undergo above room temperature
a Peierls transition mediated by an acoustical
long-wavelength and an optical $q=2k_F$ phonons respectively. In the armchair
geometry, we verify that the electron-phonon coupling  parameter $\lambda$
originates mainly from phonons at $q=2k_F$ and is strongly enhanced when the
diameter decreases. These results question the origin of superconductivity in
small diameter nanotubes.

\end{abstract}
%==================================================================

\pacs{61.46.+w, 61.50.Ah, 63.20.Kr, 63.22.+m, 68.35.Rh, 74.25.Kc, 74.70.Wz }

%61.46.+w Nanoscale materials: clusters, nanoparticles, nanotubes, and nanocrystals
%61.50.Ah Crystalline state: Theory of crystal structure, crystal symmetry; 
%         calculations and modeling
%63.20.Kr Phonons in crystal lattices: Phonon-electron and phonon-phonon interactions
%63.22.+m Phonons or vibrational states in low-dimensional structures and nanoscale materials
%68.35.Rh Phase transitions and critical phenomena
%74.25.Kc Properties of type I and type II superconductors: Phonons
%74.70.Wz Superconducting materials : Fullerenes and related materials

\maketitle

%% Intro
Superconductivity (SC) has been recently discovered in the 4-\AA\ diameter carbon
nanotubes (CNTs) embedded in a zeolite matrix \cite{ Tang01} 
with a transition temperature $T_{SC}$= 15~K,
much larger than that observed in the bundles of larger diameter tubes~\cite{
Kociak01}. This has stimulated a significant amount of work at the
theoretical level to understand the origin of the SC
transition \cite{ Gonzalez01, Sedeki02, Kamide03, Martino03}.  While both
reduced screening and quantum/thermal fluctuations disfavor
superconductivity in 1D systems, early work on fullerenes \cite{Schluter92}
and nanotubes \cite{Benedict95} suggested that the curvature significantly
enhances the strength of the electron-phonon (e-ph) coupling.  This is
consistent with the softening of the Raman modes under
tube diameter reduction~\cite{ Dubay02}.

%% Previous work

In 1D systems, the enhancement of the e-ph deformation potential should also
favor the occurrence of Peierls, or charge density wave (CDW), transitions
driving the system into an insulating phase.  This possibility has been
explored in early works on the basis of model Hamiltonians \cite{
Mintmire92}.  The strength of the e-ph coupling was either extrapolated from
its value in graphite or taken as an adjustable parameter.  The Peierls
instability transition temperatures $T_{CDW}$ were estimated to be much
smaller than room temperature: for instance, Huang {\it et al.} \cite{
Mintmire92} find $T_{CDW}$ $\sim$ 9.1 K for a C(5,5) tube.  Further,
long-wavelength optical \cite{ Dubay02} or acoustical \cite{ Figge01}
``opening the gap'' phonon distortions have also been explored and the
transition temperature was again predicted to be a few degrees Kelvin.

The Peierls transition in 4-\AA\ diameter CNTs has not been addressed
explicitly, except for a recent tight-binding exploration of the C(5,0)
tube~\cite{ Kaxiras04} and an {\it ab initio} study of the C(3,3) one
\cite{CTChan04}. While the inclusion of electronic screening  in the
non-self-consistent approach~\cite{Kaxiras04} for the C(5,0) case yields
a negligible $T_{CDW}$, the {\it ab initio} simulations predict for the
C(3,3) tube a $T_{CDW}$ temperature of $\sim$ 240~K. These results suggest
that the C(5,0) tube is the only candidate for a SC transition at 15~K.
However, such a large discrepancy between the two results call for a
unified treatment of both nanotubes within a parameter-free approach.

In the present work, we study by means of {\it ab initio} simulations the
phonon band structure and e-ph coupling in 4-\AA\ diameter CNTs. We
show that the C(5,0) and C(3,3) tubes
both undergo a Peierls distortion with a critical
temperature $T_{CDW}$ larger than room temperature. While the distortion is
associated with a long-wavelength acoustical mode in the C(5,0) case, the
e-ph coupling is dominated by phonons at $q=2k_F$ in the
case of C(n,n) nanotubes with a coupling parameter $\lambda$ which is
strongly enhanced with decreasing diameter.

The electronic and vibrational properties are studied within a
pseudopotential~\cite{ Troullier91} planewave approach to density functional
theory (DFT). We adopt the local density approximation (LDA) and a 50 Rydb
cut-off to expand the electronic eigenstates.  The phonon modes are
calculated at arbitrary $q$-vectors using the perturbative DFT
approach~\cite{ Baroni01} (in what follows $q$- and $k$-vectors label phonon
and electron momenta respectively).  Special care is taken in sampling the
electronic states around the Fermi level using an adaptative $k$-grid
technique. The energy levels are populated using either a
Gaussian broadening \cite{ Methfessel89} or a Fermi-Dirac (FD) distribution
within Mermin's generalization \cite{ Mermin65} of DFT to the canonical
ensemble. The latter technique explicitly accounts for the effect of
temperature. We use a periodic cell allowing for 10 \AA\ vacuum between
the nanotubes.

\begin{figure}
\begin{center}
\includegraphics[width=7cm]{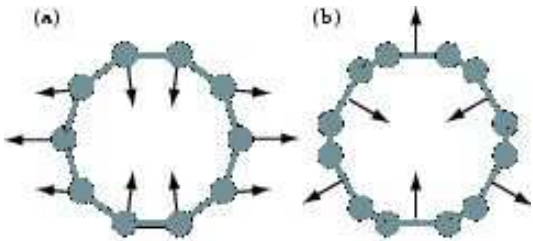}
\caption{Symbolic representations of the out-of-plane acoustical and optical
modes at $\Gamma$ for the phonon bands driving the Peierls instability in the
(a) C(5,0) and (b) C(3,3) tubes, respectively. }
\label{fig:phonmode}
\end{center}
\end{figure}

We start by studying the nanotubes band structure after careful structural
relaxation with a 25~meV FD distribution broadening (T$\sim$300~K).
The first important outcome of this calculation is the spontaneous
zone-center deformation of the C(5,0) tube~\cite{ note:deform}. From the
D$_{10h}$ symmetry, the structure relaxes to form an elliptic tube (D$_{2h}$
symmetry), as indicated by the arrows in Fig.~\ref{fig:phonmode}(a), with
principal axes of 3.83 \AA\ and 4.13 \AA.

\begin{figure}
\begin{center}
\includegraphics[width=7cm]{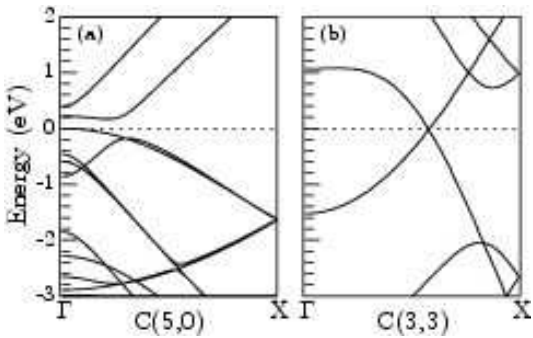}
\caption{Band structures of (a) the ``distorted'' D$_{2h}$ C(5,0) nanotube
and (b) the C(3,3) tube.  The zero of energy has been set to the top of the
valence bands and at the Fermi level, respectively.
}
\label{fig:elecband}
\end{center}
\end{figure}

The corresponding band structure is provided in Fig.~\ref{fig:elecband}(a).
As a result of the reduction of symmetry, the repulsion of the hybridized
$\sigma$-$\pi$ bands~\cite{ Blase93} at the Fermi level opens a $\sim$ 0.2
eV band gap (LDA value). This implies that {\it the C(5,0) nanotube undergoes
a metallic-semiconducting transition with T$_{CDW}$ larger than room
temperature}. Following Ref.~\onlinecite{ Dubay02}, such a transformation
can be referred to as a ``pseudo-Peierls'' transition since it opens the band
gap through a (long-wavelength) structural deformation. 

In the C(3,3) case, we do not observe any zone-center
deformation leading the band structure away from the well-documented \cite{
Machon02} zone-folding picture of two linear bands crossing at the Fermi
level $E_F$.  However, compared to the zone-folding analysis, the Fermi
wavevector $k_F$ is no longer at 2/3 $\Gamma$X but at $\sim$ 0.58 $\Gamma$X so
that $2\pi/2k_F$ is no longer commensurate with the unperturbed lattice (or
commensurate with a very large supercell). This is a well documented effect
of the curvature.

\begin{figure}
\begin{center}
\includegraphics[width=7cm]{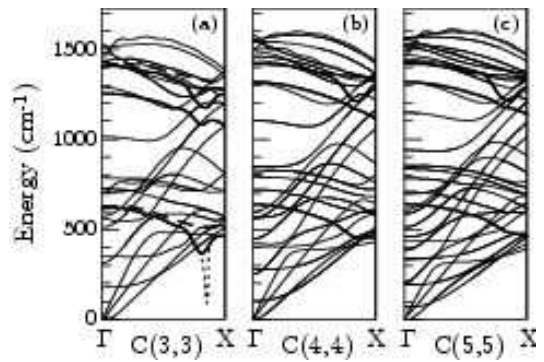}
\caption{Phonon band structures for the C(n,n) nanotubes with n=3, 4, and 5
from left to right obtained with a Gaussian broadening of 136~meV (10~mRy).
The dotted line for the C(3,3) nanotube corresponds to the additional
explicit calculation of the phonon modes around $q=2k_F$ with a reduced
68~meV (5~mRy) broadening. Important bands appear as thicker lines.  The
``undressed'' $\omega^0_{q}$ bands (see text) are indicated with thick
dashed lines.  }
\label{fig:phonband}
\end{center}
\end{figure}

We further present in Fig.~\ref{fig:phonband}(a) the phonon band structure of
the C(3,3) tube~\cite{ note:phonband}.  The most salient feature is {\it the
giant Kohn anomaly \cite{ Kohn59} at $q=2k_F$ leading to a dramatic softening
of a few phonon modes}~\cite{ note:classical}. In particular, the optical
band starting at 620 cm$^{-1}$ at $\Gamma$ becomes the lowest vibrational
state at $q=2k_F$ [thick/dotted low lying band in
Fig.~\ref{fig:phonband}(a)]. The associated atomic displacements, 
indicated in Fig.~\ref{fig:phonmode}(b), correspond to an out-of-plane
optical mode in agreement with the results of Ref.~\onlinecite{ Kaxiras04}.
This is consistent with the analysis provided by Mahan \cite{ Mahan03}: the
coupling of electrons with out-of-plane optical modes should be enhanced away
from zone-center and with decreasing diameter.  The predominance of such a
mode for e-ph coupling is a strong manifestation of the importance of
curvature in small diameter tubes.

The value of the associated frequency at $q=2k_F$ strongly depends
on the energy distribution broadening used in the  phonon
calculation, as clearly illustrated in Fig.~\ref{fig:phonband}(a) by the
difference between the thick and dotted lines.  Using a 25~meV FD
distribution (T$\sim$300~K), we find a negative phonon mode
$\omega_{2k_F,\nu=1}$ of -200 cm$^{-1}$.  This means that {\it the C(3,3)
tube undergoes a Peierls distortion with T$_{CDW}$ larger than room
temperature \cite{ note:bracketing}. Consequently, at the mean-field SC 
transition temperature measured for the 4-\AA\ diameter CNTs ($T_{SC}$=15~K), 
the C(5,0) and C(3,3) tubes should both be semiconducting, 
seriously questioning the origin of superconductivity in small 
diameter tubes.} 

Our results are in good agreement with those of Ref.~\onlinecite{CTChan04}
concerning the C(3,3) tube, but contrast significantly 
with those based on a model for the
interaction Hamiltonians and the electronic susceptibility~\cite{ Kaxiras04}.
We emphasize that our self-consistent
treatment automatically includes the renormalization of both the e-ph and e-e
interactions by electronic screening at the DFT level. 
%%The discrepancies can
%%certainly be explained by the exponential dependence of the transition
%%temperature on the interaction parameters (namely, $\lambda$ and $\mu$).

%%
To study the influence of tube diameter on the softening of the modes at
$q=2k_F$, we represent in Fig.~\ref{fig:phonband}(b,c) the phonon band
structure of the C(n,n) tubes with n=4 and 5 \cite{ note:phonband}. Again, we
observe that the same optical band (thick low lying band) is softened at
$2k_F$, but clearly the effect is strongly reduced with increasing diameter.
This confirms that T$_{CDW}$ will quickly lower with increasing diameter.
The large amount of {\it sp}$^3$ character in ultra-small tubes invite
to draw a comparison with the SC transition in doped carbon
clathrates~\cite{clathrates} and diamond~\cite{diamond}.

Surprisingly, the softening of another mode at higher energy (starting as an
optical mode around 1450 $cm^{-1}$, also indicated by a thick line) seems to
be less sensitive to the diameter. This softening is the signature of a
large e-ph coupling which might thus play an important 
role in the transport properties of larger CNTs.~\cite{mauri04}

To further quantify the strength of the e-ph interaction, we can calculate
the e-ph coupling matrix elements $g_{q\nu}(knn')$ and the coupling constant
$\lambda$:
\begin{eqnarray}
& & \lambda =  \sum_{q,\nu} \lambda_{q,\nu} = 2
N(E_f) \sum_{q,\nu} < |g_{q\nu}|^2> / \hbar \omega_{q\nu} \\
& & < |g_{q\nu}|^2>
= \int { \frac{dk}{L_{BZ}} } \sum_{n,n'} |g_{q\nu} (knn')|^2 {
\frac{\delta(\epsilon_{k+q,n'}) \delta(\epsilon_{kn})}{N(E_F)^2} }
\\
& & g_{q\nu}(knn') =
({\frac{\hbar}{2M\omega_{q\nu}}})^{1/2} <  \psi^0_{n{\bf k}}
| {\bf \hat \epsilon_{q\nu}} \cdot {\frac{\delta V}{\delta \hat{u}_{q\nu}} }
| \psi^0_{n'{\bf k+q}}>
\end{eqnarray}

\noindent
Due to conservation of energy and momentum, only vibrational states around
$q=\Gamma$ and $\pm 2k_F$ contribute to $\lambda$, as the nesting factor
\cite{ note:nestfact} n(q) quickly decays away from these points.  At
$\Gamma$, only four modes contribute to $\lambda$ due to symmetry
considerations. They are the radial breathing mode (RBM), the optical
out-of-plane (ZO) mode, and the in-plane optical longitudinal A$_1$(L) and
transverse A$_1$(T) (G-band) modes (see Ref.~\onlinecite{ Dubay02})
located respectively at 536 cm$^{-1}$, 692 cm$^{-1}$, 1354 cm$^{-1}$ and
1431 cm$^{-1}$ in the C(3,3) tube.

In Table~\ref{tab:coupling}, we report the parameter $\lambda(\nu,q \subset
\{\Gamma\})$ defined as the $q$-sum of all the $\lambda_{q,\nu}$ (see Eq.~1)
for $q$ in the neighborhood of $\Gamma$ on a given $\nu$-band \cite{
note:sum}.  This coupling strength varies from one band to another and is
enhanced when the tube diameter decreases. However, the variation with the
diameter differs from one mode to another and it is difficult to extract a
simple scaling law.

\begin{table}
\caption{Contributions ($\lambda_\nu$) to the coupling constant for all the
$q$-vectors in the neighborhood of $\Gamma$ $(q \subset \{\Gamma\})$ and
$2k_F$ $(q \subset \{2k_F\})$ for the C(n,n) tubes with n=3, 4, and 5. The
corresponding frequencies ($\omega_\nu$) at $\Gamma$ and $2k_F$ are indicated
in cm$^{-1}$. For the C(3,3) tube, the values underlined for $q \subset
\{2k_F\}$ correspond to the ``undressed'' frequencies (see text). The global
coupling constant ($\lambda$) is indicated at the last line.  }
\label{tab:coupling}
\begin{ruledtabular}
\begin{tabular}{rrrrrrrrr}
 &\multicolumn{2}{c}{C(3,3)} &
 &\multicolumn{2}{c}{C(4,4)} &
 &\multicolumn{2}{c}{C(5,5)} \\
 &\multicolumn{1}{c}{$\omega_\nu$}
 &\multicolumn{1}{c}{$\lambda_\nu$} &
 &\multicolumn{1}{c}{$\omega_\nu$}
 &\multicolumn{1}{c}{$\lambda_\nu$} &
 &\multicolumn{1}{c}{$\omega_\nu$}
 &\multicolumn{1}{c}{$\lambda_\nu$} \\
\hline
\multicolumn{1}{l}{$q \subset \{\Gamma\}$}\\
RBM      &  536 & 0.009 &&  416 & 0.005 &&  338 & 0.001 \\
ZO       &  692 & 0.009 &&  799 & 0.004 &&  848 & 0.001 \\
A$_1$(L) & 1354 & 0.007 && 1468 & 0.009 && 1538 & 0.002 \\
A$_1$(T) & 1431 & 0.012 && 1505 & 0.009 && 1570 & 0.002 \\
\multicolumn{1}{l}{$q \subset \{2k_F\}$}\\
         & \underline{520}&\underline{0.074}&&  414 & 0.022 &&  497 & 0.010 \\
         &            499 &           0.026 &&  509 & 0.010 &&  533 & 0.004 \\
         &\underline{1320}&\underline{0.018}&& 1013 & 0.010 && 1154 & 0.009 \\
         &           1144 &           0.001 && 1180 & 0.014 && 1237 & 0.011 \\
\hline
 & & 0.156 &
 & & 0.083 &
 & & 0.040 \\
\end{tabular}
\end{ruledtabular}
\end{table}
We now discuss the e-ph coupling at $q=2k_F$. The strong variation
of the lowest frequency $\omega_{2k_F,\nu=1}$ with temperature raises the
question of the meaning of $\lambda_{2k_F,\nu=1}$ calculated with the
$g_{q\nu}$ and $\omega_{q\nu}$ values obtained from {\it ab initio}
calculations.  The difficulties have been clearly exposed in
Ref.~\onlinecite{ Kaxiras04}.  If one is interested in studying the SC
transition, one faces the difficulty that at the experimental T$_{SC}$, the
self-consistent $g_{2k_F,\nu}$ and $\omega_{2k_F,\nu}$ values are those of
the CDW phase. As $T_{CDW}$ is larger than any realistic SC transition
temperature, we will not attempt to estimate T$_{SC}$ with e-ph vertices and
ph-propagator properly ``undressed'' from the CDW instability~\cite{
Kaxiras04}. We limit ourselves to identifying the relevant modes and
providing the values for $\lambda_{2k_F,\nu=1}$ obtained 
above $T_{CDW}$, which are those that matter for
transport measurements in the normal state.

While the $g_{2k_F}$ parameters are rather stable with temperature above
$T_{CDW}$, the phonon frequencies still vary significantly. Following
Ref.~\onlinecite{ Kaxiras04}, we provide an upperbound for
$\omega_{2k_F,\nu}$ by removing the Kohn's anomaly. Namely, we assume that
around $q=2k_F$,

\begin{equation}
\omega^2_{q,\nu} = (\omega^0_{q,\nu})^2 +
 A \times ln{\Big |}(q-2k_F)/(q+2k_F){\Big |}
\end{equation}

\noindent
and fit the $\omega^2_{q,\nu}$ bands by a cubic polynomial for
($\omega^0_{q,\nu})^2$ plus the logarithmic term that we subtract to obtain
the bare phonon frequency. The important ``undressed'' bands (corresponding
to $\omega^0_{q,\nu}$) are represented by dashed lines in
Fig.~\ref{fig:phonband}(a). The related frequencies and $\lambda_{2k_F,\nu}$
values are underlined in Table~\ref{tab:coupling}.  This treatment is applied
to both states presenting the largest Kohn anomaly, but its effect is clearly
more pronounced for the low-lying $\nu=1$ band.

Using the $\omega^0_{2k_F,\nu}$ frequencies leads to providing a lower bound
for the corresponding $\lambda_{2k_F,\nu}$ e-ph parameter. Even in that
limit, we observe that the coupling with modes at $q=2k_F$ is significantly
larger than with long-wavelength phonons. As at zone-center, few modes with
$q=2k_F$ couple to the electrons and the strength of the coupling
is significantly enhanced with decreasing diameter. In the C(3,3) case,
we obtain in the undistorted phase a density of states $N(E_F)$ = 0.4 
states/eV-cell-spin, yielding an e-ph potential $V^{ep} = \lambda/N(E_F)$ =
440~meV. This is larger than that in the fullerenes (60-70~meV),
but it is not sufficient to lead to any significant $T_{SC}$ value. 
The much larger density of states in the 
``$D_{10h}$'' C(5,0) tube ($N(E_F)$ = 1.8 states/eV-cell-spin) may lead
in principle to a large $T_{SC}$ value.  However, as shown above, 
the prevailing structure around the experimental $T_{SC}$
temperature is the insulating $D_{2h}$ one.

In conclusion, we have shown that the ``metallic'' C(5,0) and C(3,3) nanotubes
undergo a Peierls transition with a critical temperature larger than 300~K.
Our results are derived within Mermin's generalization of DFT to finite
temperature systems and do not rely on models for the e-ph interactions nor
the electronic susceptibility. Additional physical ingredients (interaction
with the zeolite network, possible defective structure, Luttinger liquid
character, etc.) might reconcile these results with the experimentally
observed SC transition at T$_{SC}$=15~K.

\noindent {\bf Acknowledgements:}
Calculations have been performed at the French CNRS national computer center
at IDRIS (Orsay). GMR and JCC acknowledge the National Fund
for Scientific Research [FNRS] of Belgium for financial support. Parts of
this work are also connected to the Belgian Program on Interuniversity
Attraction Poles (PAI5/1/1) on Quantum Size Effects in Nanostructured
Materials, to the Action de Recherche Concert\'ee ``Interaction
\'electron-phonon dans les nanostructures'' sponsored by the Communaut\'e
Fran\c caise de Belgique, and to the NanoQuanta and FAME European networks of
excellence.

%===================== Bibliography =========================


\begin{thebibliography}{10}
\bibitem{note:phonband}
The phonon band structures are obtained by a special interpolation
technique~\cite{ Baroni01} based on the explicit calculation of the dynamical
matrices on 20 $q$-point grid.
\bibitem{note:bracketing}
Using various values of FD broadening, T$_{CDW}$ can be estimated to lie with
300~K and 400~K.
\bibitem{note:nestfact}
$n(q)$ can be obtained by setting the $g$ coefficients to one in Eq.~2. It is
a measure of ``phase space availability'' for electrons to be scattered by
phonons of wavevector $q$.
\bibitem{note:sum}
The domain of integration is defined by the extent of $n(q)$ around $\Gamma$.
We have observed that $\lambda(q,\nu)/n(q)$ is nearly independent of q in
this domain. This greatly helps in calculating $\lambda$ as $n(q)$ can be
calculated with arbitrary precision by fitting the electronic bands around
$\pm$ $k_F$.
\end{thebibliography}
\end{document}